\documentstyle[12pt]{article}

\newcommand{\be}{\begin{equation}}
\newcommand{\ee}{\end{equation}}
\newcommand{\bea}{\begin{eqnarray}}
\newcommand{\eea}{\end{eqnarray}}
\newcommand{\la}{\label}
\newcommand{\nono}{\nonumber}
\newcommand{\ci}{\cite}
\newcommand{\bi}{\bibitem}
\newcommand{\E}{{\cal E}}
\newcommand{\half}{\frac{1}{2}}
\newcommand{\eq}{\rm eq}
\newcommand{\tauc}{\tau_{\rm c}}

\begin{document}

\title{{\bf Back-reaction in the presence of thermalizing 
collisions}
\vspace{1 true cm}
\author{J.M. Eisenberg \\
\\ Institut f\"ur Theoretische Physik,
Universit\"at Frankfurt \\
60054 Frankfurt-am-Main, Germany \\
\centerline{and} \\
School of Physics and Astronomy \\ 
Raymond and Beverly Sackler Faculty of Exact Sciences \\ 
\vspace {1 true pc}
Tel Aviv University, 69978 Tel Aviv, Israel\thanks{Permanent address.
Email address: judah@giulio.tau.ac.il.}}}
\date{August, 1996}

\maketitle

\centerline{\sc Dedicated to the memory of Larry Biedenharn}

\begin{abstract}
\baselineskip 1.5 pc
Preequilibrium parton production following an ultrarelativistic 
nucleus--nucleus collision is studied in terms of the decay of a strong 
chromoelectric field which generates pairs through the Schwinger
mechanism.  Back-reaction of the partons with the field is included and a
model transport equation containing a collision term is solved for 
the central rapidity region based 
on an approximation in which the partons relax to a thermal distribution.
\end{abstract}
\vfil
\newpage
\parindent 2 pc
\parskip 1 pc
\baselineskip 1.5 pc

\section{Introduction}

In recent years it has proved possible to solve the problem of
back-reaction in the context of preequilibrium parton production in the
quark--gluon plasma \ci{CEKMS,KES}.  This addresses the scenario in which
two ultrarelativistic nuclei collide and generate color
charges on each other, which in turn create a chromoelectric field
between the receding disk-like nuclei.  Parton pairs then tunnel
out of this chromoelectric field through the Schwinger mechanism
\ci{Sau,HE,Sch} and may eventually reach thermal
equilibrium if the plasma conditions pertain for a sufficient length of
time.  While the tunneling and the thermalizing collisions proceed, the
chromoelectric field accelerates the partons, producing a current which
in turn modifies the field.  This back-reaction may eventually set up 
plasma oscillations.

This picture for preequilibrium parton production has been studied
in a transport formalism \ci{BC} in which the chromoelectric field is
taken to be classical and abelian and collisions between the partons are 
completely ignored so that the only interaction of the partons is with 
the classical electric field, this interaction being the source of the
back-reaction.  Alternatively, the mutual scattering
between partons has been
considered in an approximation that assumes rapid thermalization and
treats the collisions in a relaxation approximation about the thermal
distribution \ci{KM}; in this study no back-reaction was allowed.  
Both interparton collisions and back-reaction were considered in a
calculation \ci{GKM} done in the hydrodynamic limit, which thus took 
into account only electric conduction within the parton plasma.  
All of these studies focused on the region of central rapidity.

The more recent calculations \ci{CEKMS,KES} carried out a comparison
between the transport formalism for back-reaction and the results of a
field-theory calculation for the equivalent situation (see also 
\ci{CM,KESCM1,KESCM2}) and found a remarkable similarity between the
quantal, field-theory results and those of the classical transport
equations using a Schwinger source term.  This link has also been
established formally to a certain degree \ci{BE}.  (This close
relationship tends to fail, in part, for a system 
confined to a finite volume as a dimension of this
volume becomes comparable with the reciprocal parton effective
mass \ci{Eis1}.)
The studies relating field theory with transport formalism were all
carried out under the assumption of a classical, abelian electric field
and no parton--parton scattering.  The removal of the assumption of a
classical field, and thus the inclusion of interparticle scattering
through the exchange of quanta, has been considered quite recently
\ci{CHKMPA} for one spatial dimension.

The study reported here is carried out within the framework of the transport
formalism (the parallel field-theory case is also currently under
study \ci{Eis2}) and incorporates both back-reaction and a collision term
in the approximation of relaxation to thermal equilibrium.  Thus it
assumes that thermalization takes place fast enough so that it makes
sense to speak of the ongoing tunneling of partons, with back-reaction,
as the collisions produce conditions of thermal equilibrium.  It 
may be seen as combining the features of the studies of Bia\l as and
Czy\.z \ci{BC} with those of Kajantie and Matsui \ci{KM}, or of
paralleling the calculation \ci{GKM} of Gatoff, Kerman, and Matsui, 
but at the level of the transport formalism without further appeal to
hydrodynamics.  Along with the other studies noted, it restricts itself
to the region of central rapidity.  The study provides a model for
comparing the interplay between the thermalizing effects of particle 
collisions and the plasma oscillations produced by back-reaction.

\section{Formalism}

The transport formalism for back-reaction using boost-invariant variables
has been presented previously in considerable detail \ci{CEKMS} 
and is modified here
only by the appearance of the collision term in the approximate form
appropriate to relaxation to thermal equilibrium \ci{KM}.  The
Boltzmann--Vlasov equation in $3 + 1$ dimensions then reads, in the
notation of \ci{CEKMS},
\be \la{BV}
p^\mu\frac{\partial f}{\partial q^\mu} 
- ep^\mu F_{\mu\nu} \frac{\partial f}{\partial p_\nu} = S + C,
\ee
where $f = f(q^\mu,p^\mu)$ is the distribution function, $S$ is the
Schwinger source term, and $C$ is the relaxation-approximation collision 
term.  The electromagnetic field is $F_{\mu\nu}$ and the electric charge
$e.$ The variables we take are
\be \la{variables}
q^\mu = (\tau,x,y,\eta),\quad\quad p_\mu = (p_\tau,p_x,p_y,p_\eta),
\ee
where $\tau = \sqrt{t^2-z^2}$ is the proper time and 
$\eta = \half\log[(t+z)/(t-z)]$ is the rapidity.  Thus, as usual, the
ordinary, laboratory-frame coordinates are given by
\be \la{labcoords}
z = \tau\sinh\eta, \quad\quad t = \tau\cosh\eta.
\ee
The momentum coordinates in eq.~(\ref{variables}) relate to the
laboratory momenta through
\be \la{momenta}
p_\tau = (Et - pz)/\tau, \quad\quad p_\eta = -Ez + tp,
\ee
where $E$ is the energy and $p$ is the $z$-component of the momentum, the
$z$-axis having been taken parallel to the initial nucleus--nucleus
collision direction or initial electric field direction.

Inserting expressions \ci{CEKMS,KM} for the source term $S$ and for
the collision term $C$, and restricting to $1 + 1$
dimensions, the Boltzmann--Vlasov equation becomes
\bea \la{transport}
\frac{\partial f}{\partial\tau} 
+ e\tau\E(\tau)\frac{\partial f}{\partial p_\eta} & = &
\pm(1\pm 2f)e\tau|\E(\tau)|
\log\left\{1\pm\exp\left[-\frac{\pi m^2}{|e\E(\tau)|}\right]\right\}
\delta(p_\eta)
\nono \\
& - & \frac{f - f_{\eq}}{\tauc}.
\eea
Here the upper sign refers throughout to boson production and the lower
sign to the fermion case, and we have incorporated the necessary
\ci{CEKMS} boson enhancement
and fermion blocking factor $(1\pm 2f).$  The electric field is given 
for these variables by
\be \la{E}
\E(\tau) = \frac{F_{\eta\tau}}{\tau} = -\frac{1}{\tau}\frac{dA}{d\tau},
\ee
where $A = A_\eta(\tau)$ is the only nonvanishing component of the
electromagnetic four-vector potential in these coordinates.  In
eq.~(\ref{transport}) we have assumed, as usual, that pairs emerge
with vanishing $p_\eta,$ which is the boost-invariant equivalent of the
conventional assumption that pairs are produced with zero momentum in the
laboratory frame.  

The thermal equilibrium distribution is
\be \la{feq}
f_{\eq}(p_\eta,\tau) = \frac{1}{\exp[p_\tau/T]\mp 1},
\ee
where $T$ is the system temperature, determined at each moment in proper
time from the requirement \ci{KM}
\be \la{T}
\int\frac{dp_\eta}{2\pi}\ f(p_\eta,\tau)\ p_\tau = 
\int\frac{dp_\eta}{2\pi}\ f_{\eq}[T(\tau);\, p_\eta,\tau]\ p_\tau;
\ee
here and throughout $p_\tau = \sqrt{m^2 + p_\eta^2/\tau^2},$ where $m$ 
is the parton effective mass, and the independent variables in terms of 
which the 
transport equations are evolved are $p_\eta$ and $\tau.$  In 
eq.~(\ref{transport}), $\tauc$ is the collision time or time for 
relaxation  to thermal equilibrium.

Back-reaction generates variations in $\E(\tau)$ as a function of
proper-time through the Maxwell equation
\bea \la{Maxwell}
-\tau\frac{d\E}{d\tau} = j_\eta^{\rm cond} + j_\eta^{\rm pol} & = &
2e\int\frac{dp_\eta}{2\pi\tau p_\tau}\ f\ p_\eta \nono \\
& \pm & \left[1\pm 2f(p_\eta=0,\tau)\right]
\frac{me\tau}{\pi}{\rm sign}[\E(\tau)] \nono \\
& \times & \log\left\{1\pm
\exp\left[-\frac{\pi m^2}{|e\E(\tau)|}\right]\right\};
\eea
here the two contributions on the right-hand side are for the conduction
and polarization currents, respectively.
Note that in $1 + 1$ dimensions the units of electric charge $e$ and of
the electric field $\E$ are both energy.  For numerical convenience
\ci{CEKMS} a new variable is introduced, namely,
\be \la{u}
u = \log(m\tau), \quad\quad \tau = (1/m)\exp(u).
\ee
Equations (\ref{transport}) and (\ref{Maxwell}) are to be solved as a
system of partial differential equations in the independent variables
$p_\eta$ and $\tau$ for the dependent variables $f$ and $\E,$ determining
the temperature $T$ at each proper-time step from the consistency
condition of eq.~(\ref{T}).

\section{Numerical results and conclusions}

The numerical procedures used here are patterned after those of 
ref.~\ci{Eis1}, and involve either the use of a Lax method or a method of
characteristics.  In practice the latter is considerably more efficient
in this context and all results reported here are based on it.  We note
that these methods are completely different from those used in 
ref.~\ci{CEKMS}; as a check on numerical procedures we verified that full 
agreement was achieved with the results reported there.  All quantities
having dimensions of energy are scaled \ci{CEKMS} here to units of the
parton effective mass $m,$ while quantities with dimensions of 
length are given in
terms of the inverse of this quantity, $1/m.$

In order to present a relatively limited number of cases, we fix
all our initial conditions at $u = -2$ in terms of the variable of
eq.~(\ref{u}).  At that point in proper time we take $\E = 4,$ with no
partons present; the charge is set to $e = 1.$  This has
been found \ci{CEKMS} to be a rather representative case; in particular,
little is changed by applying the initial conditions at $u = 0$ rather
than at $u = -2.$  We shall exhibit results for three values of $\tauc,$
namely 0.2, 1, and 10.  

Our results are presented in fig.~1 for boson production and in fig.~2
for fermions.  The uppermost graph in each case shows the temperature
derived from the consistency condition of eq.~(\ref{T}) while the middle
curves are for the electric field $\E$ and the lower graph gives the
total currents.  Both for bosons and for fermions, 
the cases with $\tauc = 0.2$ and $\tauc = 1$ involve a collision term 
that damps the distributions very rapidly.  Thus no signs of plasma 
oscillations, which would arise if back-reaction came into play 
unhindered, are seen.  For these values, the electric field and
total current damp rather quickly to zero, and a fixed value
of $T$ is reached.  The temperature peaks at around $1.5 m$ for the boson
cases, and near $2m$ for fermions.  The temperature ultimately achieved
depends, of course, on $\tauc.$

For $\tauc = 10,$ the plasma oscillations of back-reaction are clearly
visible in the electric field and in the total current, both for bosons
and for fermions.  In fact, these cases are rather similar to their
counterparts without thermalizing collisions \ci{CEKMS}, except for 
greater damping, especially of the current, when thermalization is 
involved.  The plasma frequency is changed only a little by this damping.
The plasma oscillations are reflected very slightly in the
temperature behavior in a ripple at the onset of the oscillations, 
where they naturally
have their largest excursion.  However, the oscillations have the effect
of pushing off the region at which a constant temperature is reached.
Extending the calculation further out in the variable $u,$ one finds that 
the temperature in that case levels off around $u \sim 5$ at a value of
$T \sim 0.38$ for bosons and 0.39 for fermions.

In conclusion, this calculation allows an exploration of the transition
between a domain dominated by parton collisions that bring about rapid
thermalization in the quark--gluon plasma and a domain governed in major
degree by back-reaction.  In the first situation, the electric field from
which the parton pairs tunnel, and the current which is produced from
these pairs by acceleration in the field, both decay smoothly to zero and
a terminal temperature is reached.  In the latter case, plasma
oscillations set in which delay somewhat the achievement of a final constant
temperature.  This qualitative difference between the two situations 
occurs for collision times about an order of magnitude larger than the
reciprocal effective parton mass.

{\sl Note added in proof:}  After this paper was completed and posted in the
Los Alamos archive, I learned of a similar study carried out by
B. Banerjee, R.S. Bhalerao, and V. Ravishankar [Phys.  Lett. B 224 (1989) 
16].  The present work has several features that are different 
from the previous one, notably, the application to bosons as well
as to fermions and the inclusion of factors for Bose--Einstein enhancement 
or Fermi--Dirac blocking in the Schwinger source term.  By comparing with the 
field theory results, these factors have been found to be of considerable 
importance \ci{CEKMS,KES,KESCM1,KESCM2}.  The earlier work uses
massless fermions, and, while the initial motion is taken to be
one-dimensional as here, it includes a transverse momentum distribution, 
so that it is difficult to make a direct quantitative comparison between 
the two studies.  There are also a number of technical differences between the
calculations.  Qualitatively, very similar behavior is found and the
earlier study points out very clearly the necessity for treating the 
interplay between back-reaction and thermalization.  I am very grateful
to Professor R.S. Bhalerao for acquainting me with this earlier reference.

It is a pleasure to acknowledge useful conversations with Fred
Cooper, Salman Habib, Emil Mottola, Sebastian Schmidt, and Ben Svetitsky
on the subject matter of this paper.  I also wish to express my warm
thanks to Professor Walter Greiner and the Institute for Theoretical
Physics at the University of Frankfurt and to Fredrick Cooper and Emil
Mottola at Los Alamos National Laboratory for their kind hospitality
while this work was being carried out.
This research was funded in part by the U.S.-Israel Binational
Science Foundation, in part by the Deutsche Forschungsgemeinschaft, and
in part by the Ne'eman Chair in Theoretical Nuclear Physics at Tel Aviv 
University.
\vfill\eject

\vskip 2 true pc

\newpage
{\Large\bf Figure captions:-}

1.  Temperature $T,$ electric field $\E,$ and total current $j$ for the
case of boson production.  The curves are labeled with the values of the
collision time or time for relaxation to thermal equilibrium $\tauc$ 
introduced in eq.~(\ref{transport}): $\tauc = 0.2,$ 1, and 10.

2.  Same as fig.~1, but for fermion production.
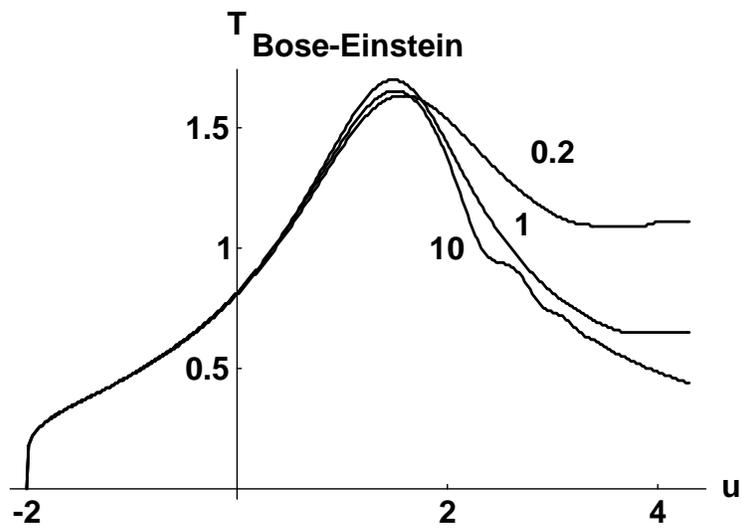
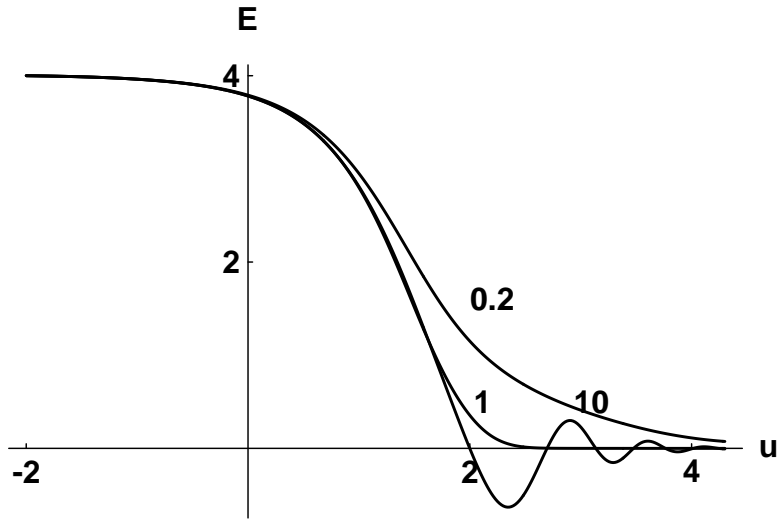
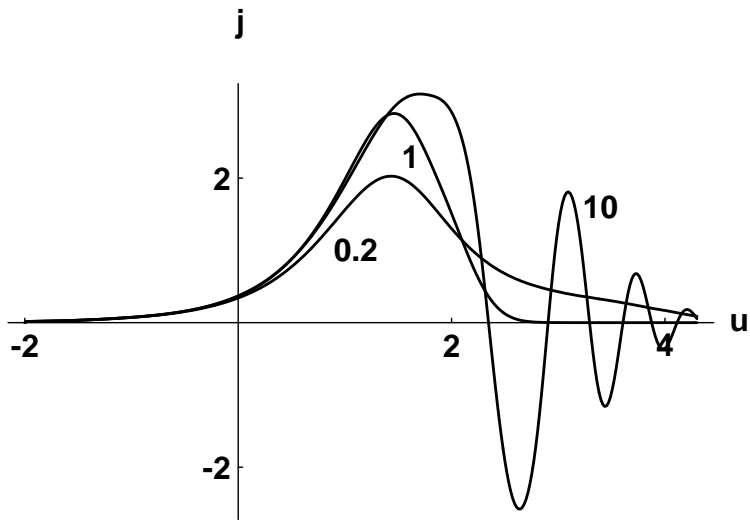
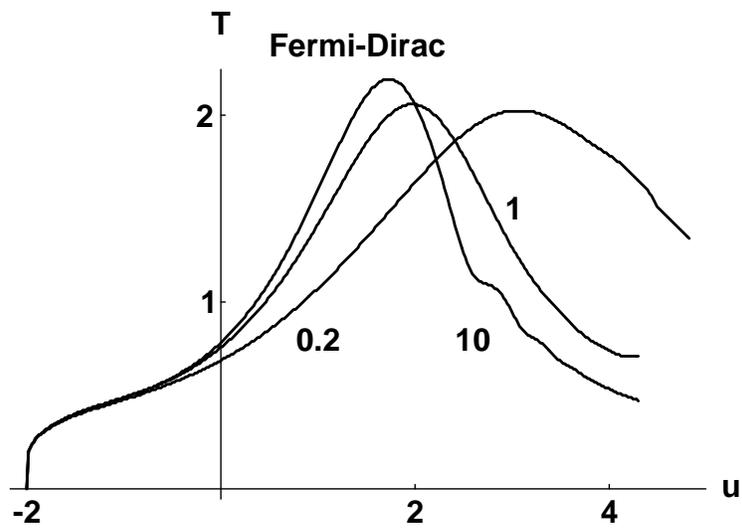
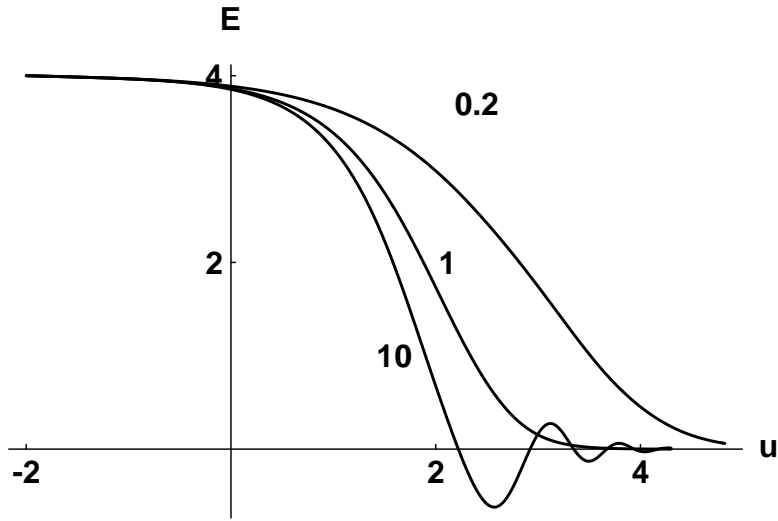
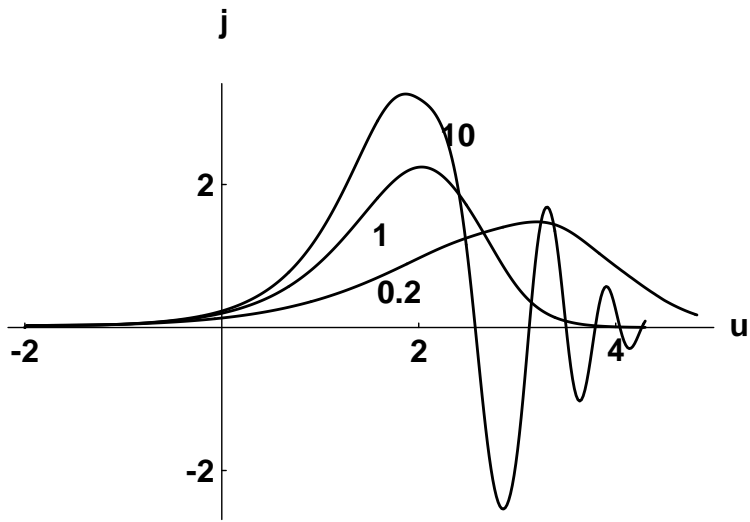
\end{document}